\documentclass[a4paper,12pt]{article}

\usepackage[colorlinks=true,linkcolor=black,anchorcolor=black,citecolor=black,filecolor=black,menucolor=black,runcolor=black,urlcolor=black]{hyperref}
\usepackage{xcolor}
\usepackage[extra]{tipa}

\usepackage{multirow,booktabs,tabularx,amssymb,colortbl,authblk,natbib}
\usepackage[T1]{fontenc}
\usepackage{mathtools}

\usepackage[margin=2.54cm]{geometry}

\newcommand{\ips}[1]{{\rmfamily $/$\textipa{#1}$/$}}

\newcolumntype{L}[1]{>{\raggedright\arraybackslash}p{#1}}
\newcolumntype{C}[1]{>{\centering\arraybackslash}p{#1}}
\newcolumntype{R}[1]{>{\raggedleft\arraybackslash}p{#1}}

\newcommand{\code}[1]{\texttt{#1}}

\begin{document}

\title{Principal components variable importance reconstruction (PC-VIR): Exploring predictive importance in multicollinear acoustic speech data}

\author[1]{Christopher Carignan}
\author[2]{Ander Egurtzegi}

\affil[1]{Speech, Hearing and Phonetic Sciences -- University College London}
\affil[2]{Centre National de la Recherche Scientifique (CNRS) -- IKER-UMR5478}

\date{} 

\maketitle

\hfill

\section{Introduction}
 
In the acoustic analysis of speech, there are a plethora of events that can be partially characterized by any number of different acoustic features. However, it is often difficult to determine which of these features are the most important for characterizing a particular speech phenomenon, especially considering the degrees of freedom introduced by researcher decisions: the choice of which acoustic variables to pursue, the techniques used to examine these variables and their effect, as well as researcher bias that may influence these decisions \citep{TimoDOF}. Additionally, the multicollinearity that will inevitably be present in a feature set designed to characterize a given speech phenomenon precludes the use of multiple regression techniques to determine variable importance \citep{Multicollinearity}. Separate regression models can be constructed for each feature in order to subsequently compare the effect sizes across models, but the threshold for determining significance must be adjusted for the multiple models (increasing the probability of Type II error) and combining the models to make predictions from new data is not straightforward.

In this paper, we present an exploratory method for determining the predictive importance of variables in multicollinear data that seeks to address these concerns. The method---principal components variable importance reconstruction (PC-VIR)---is computationally fast and simple to implement, provides results that are easy to interpret, and can be used to make predictions from new data. Within this method, the degree of variable importance is presented at two levels of significance, and techniques are described for determining importance at these two levels both with and without highly conservative adjustment for Type I error. The method is proposed here as an analytical technique to be used in exploratory analysis, not as a method of confirmatory analysis in and of itself: guided by this method as an exploratory tool, researchers can develop new, confirmatory experiments with concrete hypotheses built upon a set of statistically appropriate variables that has been derived in an objective, data-driven manner. 

\section{Methods}

\subsection{Data}

The data used in this paper to demonstrate an application of the PC-VIR method include audio recordings of six speakers of the Mixean variety of Low Navarrese Basque \citep{CaminoBasque}. This variety of Basque displays an unusual phonological contrast between oral \ips{h} and nasalized \ips{\~h}, which are vocalized in inter-vocalic position \citep{Egurt18}. For details about the data, see \citet{CaminoBasque}. The research goal of the data set for our purposes was to determine whether etymologically nasalized \ips{\~h} is still realized with phonetic nasality in the language variety. The data set was curated to include excerpts that focus specifically on this research goal, with an average of 12.3 minutes of curated speech data per speaker. The measurements created from this data set, as well as all of the functions used in the paper, can be accessed at the following address: \href{https://github.com/ChristopherCarignan/PC-VIR}{\textcolor{blue}{https://github.com/ChristopherCarignan/PC-VIR}}.

\subsection{Acoustic features}

All audio files were manually transcribed, force-aligned using the WebMAUS application \citep{WebMAUS} set for Basque (FR), and subsequently hand-corrected as needed. A set of 20 acoustic features of nasality was constructed according to the descriptions provided in \citet{Styler2017}: F1 frequency, F2 frequency, F3 frequency, F1 bandwidth, F2 bandwidth, F3 bandwidth, F1 amplitude (i.e., amplitude of the harmonic closest to F1: `A1'), F2 amplitude (i.e., amplitude of the harmonic closest to F2: `A2'), F3 amplitude (i.e., amplitude of the harmonic closest to F3: `A3'), P0 amplitude, A1-P0, P0 prominence, P1 amplitude, A1-P1, P1 prominence, P2 amplitude, A1-P2, H1-H2, spectral COG, and A3-P0. For the method demonstration shown in this paper, the goal was to determine which of these 20 acoustic features are statistically relevant for predicting the presence of vocalic nasality in the data.

Measurements of formant frequency and bandwidth, as wells as spectral COG, were obtained in Praat \citep{Praat} with speaker-optimized formant estimation parameters generated using a method similar to \citet{formantoptimization}. Measurements that rely on specific harmonics were obtained in R \citep{R} using DFT computation from the {\em wrassp} package \citep{wrassp}.

\subsection{Principal components transformation}

The 20-feature set was submitted to a separate principal components analysis (PCA) for each speaker. For each model, the principal components (PCs) with eigenvalues $\geq 1$ were retained---i.e., the Kaiser criterion of component selection\footnote{Variable selection via parallel analysis was also explored, but this approach generated the same results in these data as the (less computationally intensive) Kaiser criterion.}---resulting in an average of 5.8 PCs ($SD = 0.75$) retained for each speaker. Since the PCs are orthogonal (i.e. uncorrelated), %\footnote{Unlike with ordinary least squares regression, in which a linear fit is made by minimizing the error between the dependent variable and the model, the eigenvectors in PCA are made by minimizing the distances between the data points and their orthogonal projection to the eigenvector. This optimized linear fit ensures that the eigenvalues (i.e., the projections of the data along the eigenvector) of the first component explain the most amount of variance possible within a single dimension (i.e., the variation of the eigenvalues is maximal). Subsequently, the second eigenvector is a similarly optimized fit to the errors from the first principal component, etc., thus resulting in orthogonality between the components since the errors of each component are defined by the orthogonal projection of the data to the component's eigenvector. See \citet{Smith02} for additional background on PCA.} 
the scores for the retained PCs can be used as independent variables in subsequent statistical modeling, since there is no multicollinearity present among the variables. We demonstrate this by using the PC scores as predictors in binary logistic regression (see \citet{PCAreg} for background on PCA regression).

\subsection{Binary logistic training}

In order to map acoustic features onto the presence or absence of vocalic nasality, a binary classifier was constructed for each speaker. Training items were selected that represent unequivocally oral and nasal vocalic environments, using a variety of contextually nasalized vocalic environments and their oral vocalic environment counterparts. The contextually nasalized environments included: 10\% of the vowel interval in NVC contexts, 50\% of the vowel interval in NVN contexts, and 90\% of the vowel interval in CVN contexts. Accordingly, the oral counterparts of these environments included the same relative time points in CVC contexts. In all cases, V was one of \ips{a, e, i, o, u, y, j, w} and N was one of \ips{m, \|[n, n, \textltailn, N}. For measurements taken at the vowel midpoint (i.e. at 50\% of the vowel interval) in CVC items, both Cs were one of the stops \ips{b, p, d, t, g, k} in order to ensure that the velum was raised at the vowel midpoint. For the other two oral conditions, only the C that was closest to the relevant time point---i.e. the first C in measurements taken at 10\% of the vowel interval and the second C in measurements taken at 90\% of the vowel interval---was necessarily an oral stop. The C that was furthest from the relevant time point---i.e. the second C in measurements taken at 10\% of the vowel interval and the first C in measurements taken at 90\% of the vowel interval---was only required to be a non-\ips{h} oral consonant, and not necessarily a plosive consonant. Subsequently, the speaker-wise PC scores corresponding to the relevant time points in these oral and nasalized environments were used to construct a binomial logistic regression model with nasality (oral/nasal) as the dependent variable; nasality was dummy coded as 0 (reference: oral) and 1 (comparison: nasal). There were an average of 258.2 nasal tokens ($SD = 45.6$) and 277.7 oral tokens ($SD = 71.7$) per speaker in these models.

\subsection{Variable importance reconstruction}

The first key feature of the PC-VIR method is the ability to reconstruct the relative predictive importance of each of the original variables in a high-dimensional feature set. This reconstruction can be interpreted at three levels of statistical importance: strong importance, moderate importance, and no importance. Throughout the paper, we refer to ``importance'' in a generalized sense that can be interpreted as the strength of an independent variable in predicting the dependent variable, conflating both correlation strength and {\em z}-score effect size. We caution the reader that these levels of statistical importance are only guidelines, and we highlight again that this method should be used as an exploratory tool alone; as such, these thresholds of variable importance should not be used as confirmatory ``goalposts'' for hypothesis testing.

For each model generated using PC scores as predictor variables (e.g., for each speaker-specific oral/nasal logistic regression model in the example data used here), reconstruction of the relative importance of the original variables can be performed according to the following steps. For each of the PCs used in the model: the {\em z}-statistic in the model (i.e. the effect of the particular PC) and the loadings/coefficients for the PC (i.e. the correlations between the original variables and the particular PC) are logged. PC-weighted coefficients are then created by multiplying the {\em z}-statistic by the vector of PC loadings. Subsequently, the weighted coefficients for all of the PCs are summed for each of the original variables, resulting in a vector of importance coefficients $\vec{z'}$ for the original variables.\footnote{We would here like to note that there is a level of mathematical uncertainty regarding the use of the proposed method for scenarios that differ from those normally encountered in phonetic research on speech acoustics. The phonetician is typically concerned with understanding the relation between multiple acoustic parameters and some speech event; thus, she uses a set of parameters that have been designed specifically with the intention of characterizing (at least some aspect of) that event. Consequently, it is almost certain that many of the features will not only co-vary with the event but with each other. However, in the case of other kinds of research and associated data sets in which this co-variation is not expected, the orthogonal nature of eigenvectors in PCA may lead to destruction of information in the reconstruction of the variable importance due to, e.g., the summation of weights that include large PC loadings with opposing signs that are associated with large {\em z}-statistics with non-opposing signs.} The variable importance reconstruction is expressed as:
\begin{equation}
\displaystyle \vec{z'} = \sum_{i=1}^{n} z_i\cdot\vec{\alpha_i}
\label{eqn:coefficient}
\end{equation}
\noindent
where $n$ is the total number of PCs retained, $z_i$ is the $z$-statistic for the $i^{th}$ PC in the regression model, and $\vec{\alpha_i}$ is the vector of variable loadings for the $i^{th}$ PC. When conducting multiple analyses on the same dependent variable, it is appropriate to correct the $\alpha$ level for the number of models constructed in order to control for Type I error. Although only a single model is constructed for each speaker in the PC-VIR method, it may be prudent to control for Type I error by penalizing for the number of predictors (here, the number of PCs retained) that are included in each model \citep{Mundfrom06bonferroniadjustments}. Accordingly, $z_i$ can be Bonferroni-adjusted for $n$ in each model using the cumulative distribution function  $\Phi$ (\code{pnorm()} in R) and its inverse $\Phi^{-1}$ (\code{qnorm()} in R):
\begin{equation}
\displaystyle z_{adj} = \operatorname{sign}(z_i)\cdot\Phi^{-1}\left(1-\frac{n\cdot2\Phi(-|z_i|)}{2}\right)
\label{eqn:adjust}
\end{equation}
\noindent
where $2\Phi(-|z_i|)$ represents the two-tailed probability $Pr(>|z_i|)$; when using the functions provided in the GitHub repository, this adjustment can be implemented automatically by including \code{adjust = True} in the \code{PC\_VIR()} function. We would like to emphasize that the PCs are included as predictors in the regression models in a purely additive manner, excluding any non-linear effects. This model design preserves the linear relationships between the original variables and the importance coefficient $z'$. Models that include non-linearities may perform better in terms of predictive accuracy (see Section \ref{prediction}), but would likely perform worse in terms of inferring variable importance.

Although the $z'$ coefficient that results from the method is on the {\em z}-score scale, due to the reduction of the magnitude of the score by PC loadings with absolute magnitudes $<1$, we recommend two levels of statistical importance that take this reduction into account. The two levels of importance (moderate and strong) are defined as 0.98 and 1.372, respectively. These thresholds are derived by multiplying 1.96 (the {\em z}-statistic corresponding to a significant effect at $\alpha = 0.05$) by 0.5 and 0.7 (the conventional levels of moderate and strong correlation, respectively, in linear regression). Since the PC loadings are defined as the correlations between the original variables and the unit-scaled components (i.e. the PCs), their interpretation is similar to that in standard regression. Accordingly, these threshold values compare the product of the PC-specific variable correlations and the PC-specific {\em z}-statistics in the regression model against the product of the levels of correlation strength in linear regression and the level of significance of a {\em z}-statistic. The absolute value of a PC-VIR coefficient that meets or surpasses these levels can be interpreted as moderately or strongly important, respectively. However, as a reminder (and note of caution) to the reader: these thresholds, like all discrete cutoffs used in research, should be regarded with an appropriate degree of circumspection. By introducing two levels of statistical importance rather than a single threshold, our intention is for the method to help guide researchers' interpretation of the {\em relative} importance of individual variables, rather than providing researchers with explicit targets for confirmatory analyses.

\section{Results}

\begin{figure}[htbp!]
	\includegraphics[width=\textwidth]{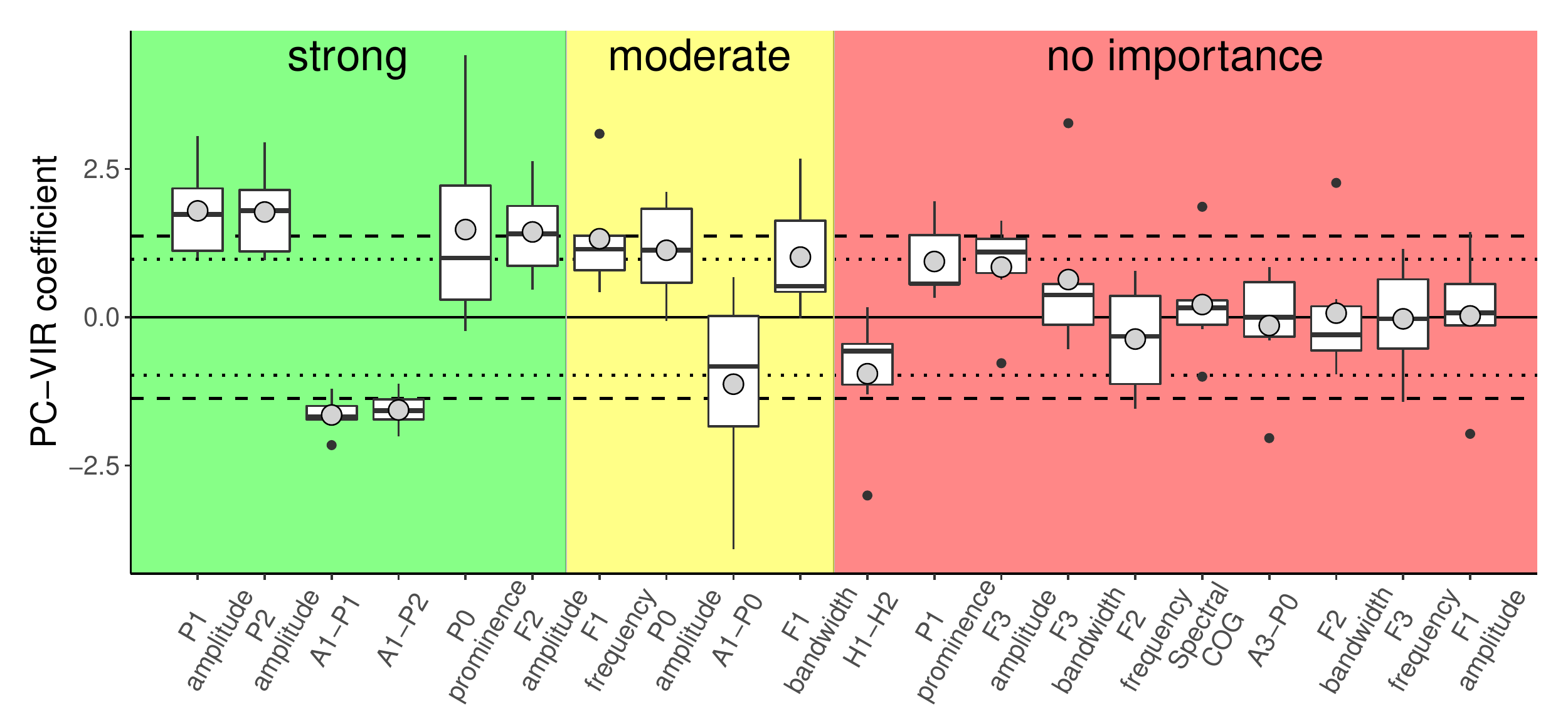}
	\caption{PC-VIR coefficients (corrected for Type I error) that predict vocalic nasality. Coefficients are displayed from left to right in descending order of the mean absolute value. Variables are grouped together as having no importance, moderate importance, or strong importance. Horizontal lines denote the levels of moderate (dotted) and strong (dashed) importance.}
    \label{figure1}
\end{figure}

Figure \ref{figure1} displays the variation of the PC-VIR coefficients for the six speaker-wise oral/nasal logistic regression models, after the correction for Type I error described in Equation \ref{eqn:adjust}. Since nasality was dummy-coded with ``oral'' as the baseline, the coefficients are interpreted with respect to their importance in distinguishing the nasal tokens from the oral reference tokens. In other words, the PC-VIR coefficients can be considered in this case as the strength and direction in the prediction of (vocalic) nasality. The inter-speaker variations for the coefficients are displayed as box plots, with the median values shown by horizontal black bars and the mean values shown by gray circles. The 20 acoustic features are displayed from left to right in descending order of the mean absolute value of the PC-VIR coefficient. Levels of moderate importance are denoted by horizontal dotted lines and levels of strong importance are denoted by horizontal dashed lines. Features with mean coefficient values that do not meet the level of moderate importance are displayed in the red portion of the plot; features that meet or surpass the level of moderate (but not strong) importance are displayed in the yellow portion; features that meet or surpass the level of strong importance are displayed in the green portion.

According to these results, the following features do not significantly predict nasality: F1 amplitude, F2 frequency, F2 bandwidth, F3 frequency, F3 bandwidth, F3 amplitude, A3-P0, spectral COG, P1 prominence, and H1-H2. Among the moderately importance features, positive coefficients are observed for F1 frequency, F1 bandwidth, and P0 amplitude: an increase in these features is associated with nasality. Conversely, a negative coefficient is observed for A1-P0: a decrease in this feature is associated with nasality. Among the strongly importance features, positive coefficients are observed for F2 amplitude, P0 prominence, P1 amplitude, and P2 amplitude, while negative coefficients are observed for A1-P1 and A1-P2.

\subsection{Method validation}

In this section, the results obtained using the PC-VIR method are verified in three ways: the direction of the effects for the original variables, how well the models generated using the PC-VIR method fit the data (i.e. goodness of fit), and how accurately values can be predicted for new data using variables selected as important by the PC-VIR method.

\subsubsection{Logistic mixed effects models}

\begin{figure}[htbp!]
\centering
	\includegraphics[width=0.8\textwidth]{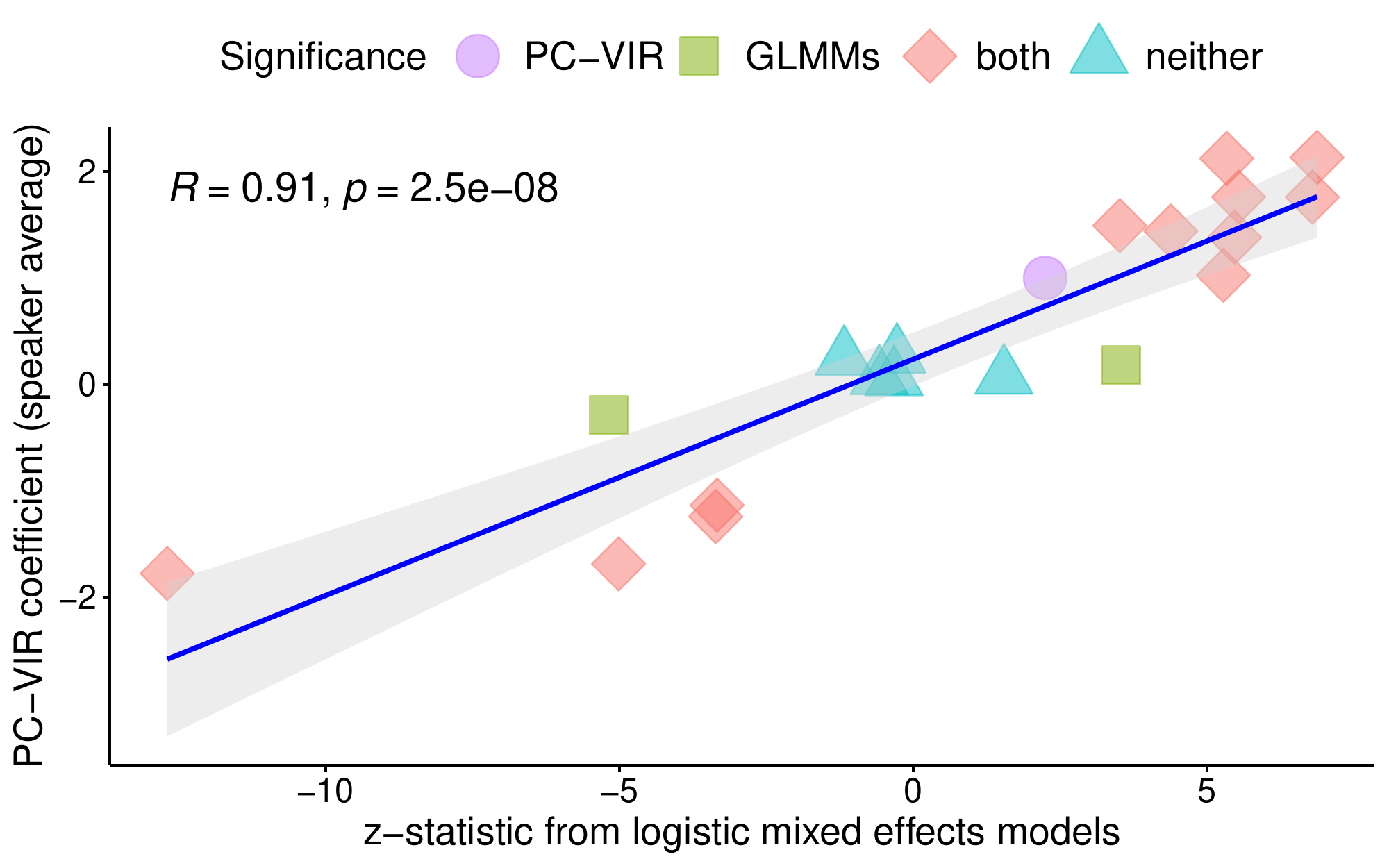}
	\caption{{\em z}-statistics from GLMMs created for each of the 20 acoustic features ({\em x}-axis) and the speaker-averaged PC-VIR coefficients ({\em y}-axis) without Type I error correction. Symbols distinguish significant effects with PC-VIR only, with GLMMs only, with both methods, and with neither. The 95\% confidence interval is indicated by the shaded band.}
    \label{figure2}
\end{figure}

\begin{figure}[htbp!]
\centering
	\includegraphics[width=0.8\textwidth]{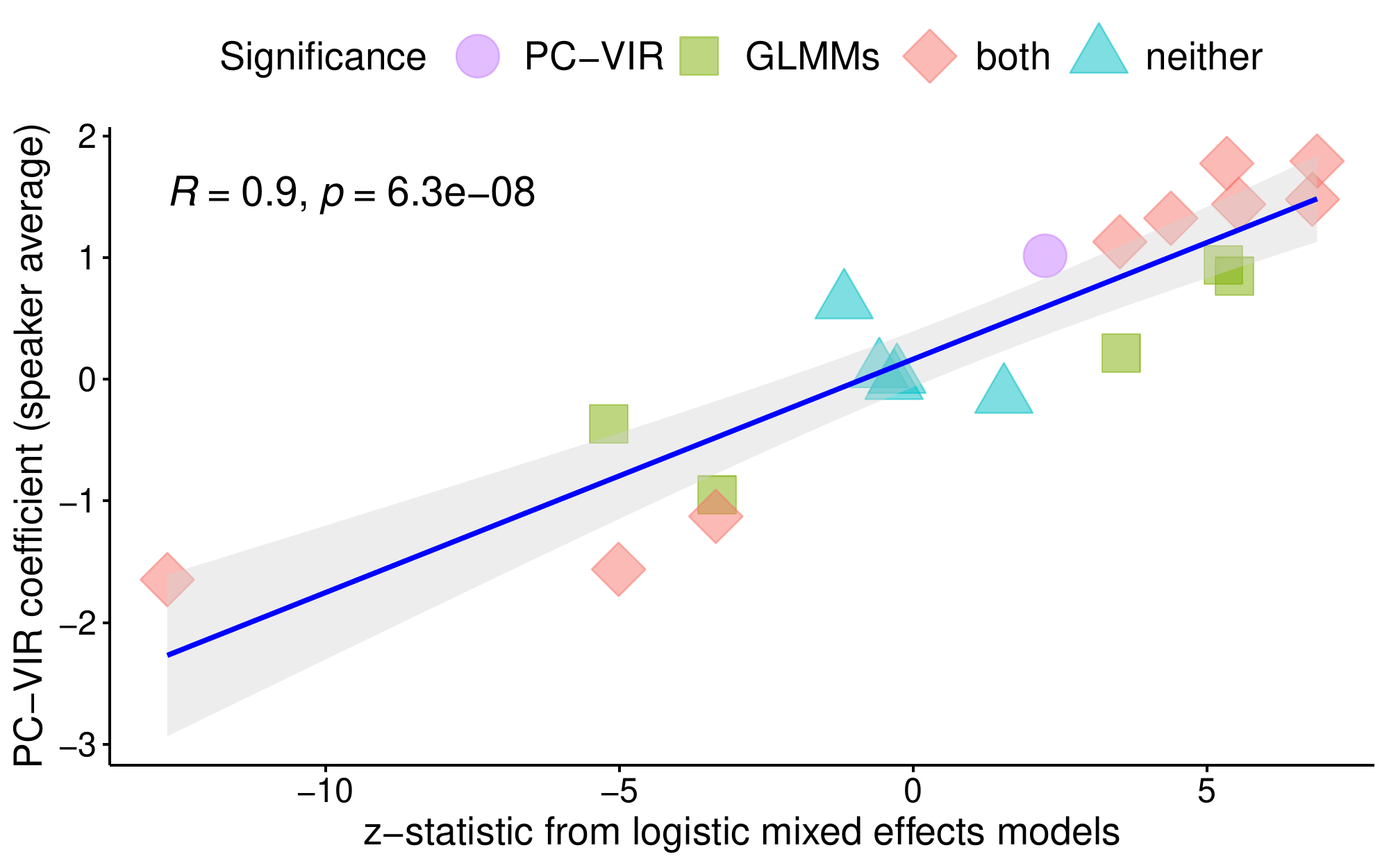}
	\caption{{\em z}-statistics from GLMMs created for each of the 20 acoustic features ({\em x}-axis) and the speaker-averaged PC-VIR coefficients ({\em y}-axis) with Type I error correction. Symbols distinguish significant effects with PC-VIR only, with GLMMs only, with both methods, and with neither. The 95\% confidence interval is indicated by the shaded band.}
    \label{figure3}
\end{figure}

The first (and arguably most important) type of verification is to determine whether the PC-VIR coefficients accurately capture the direction of effects for the independent variables. Comparison of the direction of variable effects was carried out by creating separate generalized linear mixed models (GLMMs) with logistic linking for each of the 20 acoustic features; GLMMs were created using the {\em lme4} R package \citep{lme4}. In each of the logistic GLMMs, nasality was the dependent variable, the acoustic feature was the predictor variable, random slopes and intercepts were included by phone, and random intercepts were included by speaker.\footnote{Issues with model convergence prohibited the inclusion of random slopes by speaker.} The alpha level was corrected for comparison across the 20 models using Bonferroni adjustment ($\alpha = 0.0025$). The {\em z}-statistics from the separate GLMMs are shown in comparison to both the standard PC-VIR coefficients (Figure \ref{figure2}) and the PC-VIR coefficients that include the highly conservative Type I error correction (Figure \ref{figure3}). In these figures, (purple) circles denote variables that are identified as significant with the PC-VIR method (but not in the GLMMs), (green) squares denote those that are identified as significant in the GLMMs (but not with the PC-VIR method), (red) diamonds denote those that are identified as significant with both methods, and (blue) triangles denote those that are not identified as significant with either method.

Both with and without Type I error correction, the PC-VIR coefficients are strongly correlated with the {\em z}-statistics from the separate GLMMs, revealing that the PC-VIR coefficients capture the same direction of effects as the GLMMs. However, the standard method of variable importance reconstruction (without Type I error correction) produces results that are more consistent with the results from the separate GLMMs. Without the error correction, one variable (F1 bandwidth) is significant according to the PC-VIR method but not the GLMMs, and two variables (F2 frequency and spectral COG) are significant according to the GLMMs but not the PC-VIR method. However, when including the correction for Type I error, three additional variables that are significant in the GLMMs do not reach the level of moderate importance in the PC-VIR models: H1-H2, P1 prominence, and F3 amplitude. This suggests that the method of correcting for the number of predictors in the PC-VIR models may be too conservative, inflating the overall Type II error rate.

\subsubsection{Goodness of fit}\label{fit}

Further verification was carried out by comparing goodness of fit and prediction accuracy against a well-known method that is conceptually similar to the application of logistic regression in PC-VIR: partial least squares logistic regression, which we will refer to somewhat unconventionally as ``PLS-LR'' for conciseness. PLS-LR models were created using the {\em plsRglm} R package \citep{plsRglm}. In order to compare the two methods, each speaker's data was split into training (80\%) and testing (20\%) sets via random item selection. Goodness of fit was determined using the training data sets ($\mu_N = 428.7, SD_N = 86.1$) and prediction accuracy was determined using the testing data sets ($\mu_N = 107.2, SD_N = 21.8$). Procedures were repeated 20 times using different random seeds for data splitting, and the results outlined below were similar in each case.

For both the PC-VIR and PLS-LR methods, variables identified as significantly important in the training data were used to build new PCA logistic regression models for each speaker. For the PC-VIR method, coefficients were adjusted for Type I error and acoustic features that met the threshold for moderate importance were retained. However, variable selection for the PLS-LR method is less straightforward than for the PC-VIR method. When using partial least squares linear regression, a common practice is to select variables by calculating Variable Importance in (the) Projection (VIP) scores; VIP scores $\geq 1$ are generally considered to be important to the model. However, to our knowledge, there is no current implementation of VIP score calculation for partial least squares {\em logistic} regression. Partial least squares regression (both linear and logistic) includes transformation of the data into orthogonal components, in a manner similar to PCA; with the \code{plsRglm()} function, the number of components retained is selected automatically. It is possible with this function to output whether individual variables are significant for each of the PLS components at $\alpha = 0.05$, although the method of obtaining these results is opaque to the user. For the purpose of determining goodness of fit and prediction accuracy in the data used here, variables that were significant for at least one component were retained.

After selecting important variables with both methods, new PCA logistic regression models were created for each speaker's entire data set (i.e. training and testing data combined), using only the selected variables; all PCs were retained, since the purpose of PCA-transformation in this case is for de-correlating but not reducing the data (since the data were already reduced via variable selection). The goodness of fit of the models generated by both methods was determined using Hosmer-Lemeshow (HL) tests with the {\em ResourceSelection} R package \citep{ResSel}. HL tests yield a chi-squared ($\chi^2$) value, which can be used to generate a {\em p}-value: high {\em p}-values indicate a good fit of the model to the data, while low {\em p}-values indicate a poor fitting model. For both methods, separate HL tests were created for each speaker's PCA logistic regression model, the $\chi^2$ values were averaged across the six speakers, and the averaged $\chi^2$ values were used to generate {\em p}-values for both methods using the \code{pchisq()} function. The PC-VIR method resulted in $p = 0.692$ and the PLS-LR method resulted in $p = 0.585$, indicating that both the PC-VIR and PLS-LR methods resulted in models that fit the data well.\footnote{The range of $\chi^2$ values across speakers was 3.11--14.14 ($\mu = 9.4, SD = 4.0$) for the PC-VIR method and 1.92--18.90 ($\mu = 8.19, SD = 6.13$) for the PLS-LR method. The \code{pchisq()} function suggested well-fitted models obtained from both methods in all cases except for the speaker with the lowest $\chi^2$ value obtained using the PLS-LR method ($= 1.92$).}

\subsubsection{Prediction accuracy}\label{prediction}

The second key feature of the PC-VIR method is that predictions can be made for new data, which is not easily accomplished with, e.g., the combination of separate regression models. However, it is of course crucial to determine how accurate these predictions are. Prediction accuracies for both the PC-VIR and PLS-LR methods were determined by using the PCA logistic regression models described in Section \ref{fit} to predict response values for each speaker's test data set. Values between 0 and 0.5 were coded as ``oral'' and values between 0.5 and 1 were coded as ``nasal''; the average hit rates for each method were used as a measurement of accuracy. The PLS-LR method yielded an average accuracy of 67.3\% ($SD = 1.8\%$), while the PC-VIR method yielded an average accuracy of 66.9\% ($SD = 1.9\%$) without Type I error correction and 65.9\% ($SD = 2.0\%$) with the correction, suggesting that both methods yield predictive accuracy that is well above chance. Welch two-sample {\em t}-tests revealed no difference between the two methods without Type I error correction in the PC-VIR method ($t(37.9) = -0.56, p = 0.55$), but a marginally significant difference when employing the correction ($t(37.5) = -2.19, p = 0.03$). This suggests again that correcting for the number of PCs in the PC-VIR models may be too conservative, leading to a decrease in predictive accuracy.

\section{Discussion}

The PC-VIR method that has been described in this paper is computationally fast and easy to implement, its results are easy to interpret, it is not affected by multicollinearity (cf. multiple regression), it can be used to predict new values (cf. separate regression models), and it simplifies selection of important variables (cf. PLS-LR). The PC-VIR method provides several benefits for research involving multicollinear sets of acoustic features: (1) variable importance can be interpreted at two different levels, rather than a single threshold; (2) using these levels of importance, variable selection is simple (and more straightforward than with PLS-LR); and (3) variables selected by PC-VIR result in predictions for new data that are as accurate as with PLS-LR.

\section*{Acknowledgments}
This research was funded by the Institut f\"ur Phonetik und Sprachverarbeitung (Ludwig-Maximilians-Universit\"at M\"unchen), the Alexander von Humboldt Foundation, and the Spanish Ministry of Economy and Competitiveness (FFI2016-76032-P; FFI2015-63981-C3-2). The authors gratefully acknowledge Raphael Winkelmann for his technical guidance and Michele Gubian for his mathematical insight. This paper was originally submitted for consideration for publication in {\em Journal of Acoustical Society of America -- Express Letters}; the authors are grateful to Timo Roettger and two anonymous reviewers for their careful handling of that submission process, during which their recommendations helped considerably to shape the current version of the paper.

\newpage

%=======================================================
\bibliographystyle{apalike}
\

\end{document}